\documentclass[epj]{svjour}
\usepackage{graphicx}
\usepackage{graphics}
\begin{document}
\title{Electromagnetic form factors of the baryon octet 
in the perturbative chiral quark model}
\author{S.~Cheedket\inst{1,2} \and V.~E.~Lyubovitskij\inst{1} 
Th.~Gutsche\inst{1}\and
Amand~Faessler\inst{1}\and \\K.~Pumsa-ard\inst{1}\and
Y.~Yan\inst{2}}
%
\institute{Institut f\"ur Theoretische Physik, Universit\"at
T\"ubingen, Auf der Morgenstelle 14, D-72076 T\"ubingen, Germany  
\and School of Physics, Institute of Science, 
Suranaree University of Technology, Nakhon Ratchasima 30000, Thailand}
\date{Received: date / Revised version: date}
%
\abstract{
We apply the perturbative chiral quark model at one loop to
analyze the electromagnetic form factors of the baryon octet. The
analytic expressions for baryon form factors, which are given in
terms of fundamental parameters of low-energy pion-nucleon
physics (weak pion decay constant, axial nucleon coupling, strong
pion-nucleon form factor), and the numerical results for baryon
magnetic moments, charge and magnetic radii are presented. Our
results are in good agreement with experimental data.
} 
\maketitle
\section{Introduction} 
\label{sec:intro} 
The study of the electromagnetic form factors of baryons is a very
important first step in understanding their internal structure. At
present, electromagnetic form factors and related properties
(magnetic moments, charge and magnetic radii) of the nucleon have
been measured precisely, but for the hyperons  data rarely exist
with the exception of the magnetic moments. Recently, the charge
radius of the $\Sigma^-$ has been measured~\cite{sigm01,sigm99}
and therefore gives a first estimate of the charge form factor of
the hyperon at low momentum transfers.

In Refs.~\cite{lgf01}-\cite{plgfc02} we developed the perturbative
chiral quark model (PCQM) for the study of baryon properties:
electromagnetic form factors of the nucleon, low-energy
meson-baryon scattering and $\sigma$-terms, electromagnetic
excitation of nucleon resonances, etc. In Ref. \cite{lgf01} the
PCQM has been applied to study the electromagnetic form factors of
the nucleon and the results obtained are in good agreement with
experimental data. In this paper we extend the PCQM to study the
electromagnetic form factors of hyperons and give predictions with
respect to future measurements of their magnetic moments, radii
and the momentum dependence of form factors. We proceed as
follows. In Sect.~\ref{sec:pcqm} we describe the basic notions of
our approach. In Sect.~\ref{sec:emff} we present the analytic
expressions for the charge and magnetic form factors of the baryon
octet. Numerical results for their magnetic moments, charge and
magnetic radii, and the momentum dependence of the form factors
are discussed in Sect.~\ref{sec:result}. Sect.~\ref{sec:sum}
contains a summary.

\section{
The Perturbative Chiral Quark Model}
\label{sec:pcqm}
\subsection{Effective Lagrangian and zeroth order properties}
\label{subs:leff}
The following considerations are based on the perturbative chiral
quark model (PCQM)~\cite{lgf01,lgfd01}. The PCQM is a relativistic
quark model which is based on an effective Lagrangian 
${\mathcal L}_{\rm eff} = {\mathcal L}_{\rm inv}^{lin} +{\mathcal L}_{\chi
SB}$. The Lagrangian includes the linearized chiral invariant term
${\mathcal L}_{\rm inv}^{lin}$ and a mass term ${\mathcal L}_{\chi
SB}$ which
explicitly breaks chiral symmetry 
\begin{eqnarray}\label{eq:efflg1}
{\mathcal L}_{\rm inv}^{lin}&=&\bar\psi(x)\left[i\!\!\!\not{\partial
}-\gamma^0V(r)-S(r)\right]\psi(x)\nonumber \\
&+&\frac{1}{2} \sum\limits_{i=1}^{8} [\partial_\mu \Phi_i(x)]^2  
-\bar{\psi}(x)S(r)i\gamma^5\frac{\hat{\Phi}(x)}{F}\psi(x),\\
\label{eq:efflg2}
{\mathcal L}_{\chi SB}&=& -\bar{\psi}(x){\mathcal M}
\psi(x)-\frac{B}{2}Tr[{\hat{\Phi}}^2(x){\mathcal M}],
\end{eqnarray}
where $r=|\vec x|$; $\psi$ is the quark field;
$\hat{\Phi}$ is the matrix of the pseudoscalar mesons; $S(r)$ and
$V(r)$ are scalar and vector components of an effective, static
potential providing quark confinement; ${\mathcal M}= {\rm
diag}\{\hat m,\hat m,m_s\}$ is the mass matrix of current quarks
(we restrict to the isospin symmetry limit with $m_u = m_d = \hat
m$); $B$ is the quark condensate parameter; and $F$ = 88~MeV is
the pion decay constant in the chiral limit. We rely on the
standard picture of chiral symmetry breaking and for the masses of
pseudoscalar mesons we use the leading term in chiral expansion
(i.e. linear in the current quark mass): $M^2_\pi=2\hat{m}B$,
$M^2_K=(\hat{m}+m_s)B$, $M^2_\eta=\frac{2}{3}(\hat{m}+2m_s)B$.
Meson masses satisfy the Gell-Mann-Oakes-Renner and the
Gell-Mann-Okubo relation $3M^2_\eta+M^2_\pi=4M^2_K$. In the
evaluation we use the following set of QCD parameters: $\hat m$ =
7~MeV, $m_s/\hat m = 25$ and B = $M^2_{\pi^+}/(2\hat m)$ =
1.4~GeV.

To describe the properties of baryons which are modelled as bound
states of valence quarks surrounded by a meson cloud we formulate
perturbation theory. In our approach the mass (energy)
$m^{core}_N$ of the three-quark core of the nucleon is related to
the single quark energy ${\mathcal E}_0$ by 
$m^{core}_N=3{\mathcal E}_0$. For the unperturbed 
three-quark state we introduce the
notation $|\phi_0>$ with the appropriate normalization
$<\phi_0|\phi_0> = 1$. The single quark ground state energy
${\mathcal E}_0$ and wave function (WF), $u_0(\vec x)$ are obtained
from the Dirac equation
\begin{eqnarray}
\left[-i\vec\alpha\cdot\vec\nabla+\beta S(r)+V(r)-{\mathcal E}_0\right]
u_0(\vec x) = 0. \label{eq:dirac}
\end{eqnarray}
The quark WF $u_0(\vec x)$ belongs to the basis of potential
eigenstates (including excited quark and antiquark solutions) used
for expansion of the quark field operator $\psi(x)$. Here we
restrict the expansion to the ground state contribution with
$\psi(x)= b_0u_0(\vec x)\exp(-i{\mathcal E}_0 t)$, where $b_0$ is
the corresponding single quark annihilation operator. In
Eq.~(\ref{eq:dirac}) the current quark mass is not included to
simplify our calculational technique. Instead, we consider the
quark mass term as a small perturbation.

For a given form of the potentials $S(r)$ and $V(r)$ the Dirac
equation in Eq.~(\ref{eq:dirac}) can be solved numerically. Here,
for the sake of simplicity, we use a variational Gaussian ansatz
for the quark wave function given by the analytical form:
\begin{eqnarray}\label{u_0}
u_0(\vec x) = N\exp\left[-\frac{{\vec x}^2}{2R^2}\right]\left(\begin{array}{c}
  1 \\
  i\rho\frac{\vec\sigma\cdot\vec x}{R} \\
\end{array}\right)\chi_s\chi_f\chi_c,\label{eq:u0}
\end{eqnarray}
where $N=[\pi^{3/2}R^3(1+3\rho^2/2)]^{-1/2}$ is a constant fixed
by the normalization condition $\int d^3x u_0^\dag(\vec x)u_0(\vec
x)\equiv 1$; $\chi_s$, $\chi_f$, $\chi_c$ are the spin, flavor and
color quark wave functions, respectively. Our Gaussian ansatz
contains two model parameters: the dimensional parameter R and the
dimensionless parameter $\rho$. The parameter $\rho$ can be
related to the axial coupling constant $g_A$ calculated in
zeroth-order (or 3q-core) approximation:
\begin{eqnarray}
g_A=\frac{5}{3}\left(1-\frac{2\rho^2}{1+\frac{3}{2}\rho^2}\right)=
\frac{5}{3}\frac{1+2\gamma}{3},\label{eq:ga}
\end{eqnarray}
where $\gamma=9g_A/10-1/2$. The parameter R can be physically
understood as the mean radius of the three-quark core and is
related to the charge radius $\left<r^2_E\right>^p_{LO}$ of the
proton in the leading-order (LO) approximation as
\begin{eqnarray}
\left<r^2_E\right>^p_{LO}=\frac{3R^2}{2}
\frac{1+\frac{5}{2}\rho^2}{1+\frac{3}{2}\rho^2}=
R^2\left(2-\frac{\gamma}{2}\right).\label{eq:r2p}
\end{eqnarray}
In our calculations we use the value $g_A$ = 1.25. We therefore
have only one free parameter, that is R. In the final numerical
evaluation R is varied in the region from 0.55~fm to 0.65~fm,
which is sufficiently large to justify perturbation theory.

In the PCQM confinement is introduced as a static mean field potential,
hence covariance cannot be fulfilled.
As a consequence matrix elements are frame dependent: both Galilei
invariance of the zeroth order baryon wave functions and Lorentz boost effects,
when considering finite momenta transfers, are neglected.
Approximate techniques \cite{birse,luthomas} have been developed to
account for these deficiencies in static potential models. However,
these techniques do not always agree and lead to further ambiguities in model
evaluations. Furthermore, existing Galilean projection techniques are known
to lead to conflicts with chiral symmetry constraints \cite{lgfd01}.
In the present manuscript we completely neglect the study of these additional
model dependent effects. We focus on the role of meson loops, which, as was
shown in the context of the cloudy bag model \cite{luthomas},
are not plagued by these additional uncertainties.

\subsection{Renormalization of the PCQM 
and perturbation theory}
\label{subs:rnm}
We consider perturbation theory up to one meson loop and up to
terms linear in the current quark mass. The formalism utilizes a
renormalization technique, which, by introduction  of
counterterms, effectively reduces the number of Feynman diagrams
to be evaluated. For details of this technique we refer to the
Ref.~\cite{lgf01}. Here we briefly describe the basic ingredients
relevant for the further discussion. We define the renormalized
current quark masses, $\hat m^r$ and $m^r_s$ and the
renormalization constants, $\hat Z$ and $Z_s$ as :
\begin{eqnarray}
\hat{m}^r &=& \hat{m} -\frac{3}{400\gamma}\left(\frac{g_A}{\pi
F}\right)^2\int^\infty_0 dpp^4F_{\pi NN}(p^2)\nonumber\\
&\times&\left\{\frac{9}{w^2_\pi(p^2)}+\frac{6}{w^2_K(p^2)}
+\frac{1}{w^2_\eta(p^2)}\right\},\label{eq:mhr} \\
m^r_s &=& m_s -\frac{3}{400\gamma}\left(\frac{g_A}{\pi
F}\right)^2\int^\infty_0 dpp^4F_{\pi
NN}(p^2)\nonumber\\
&\times&\left\{\frac{12}{w^2_K(p^2)}+\frac{4}{w^2_\eta(p^2)}
\right\},\label{eq:msr} \\
\hat{Z} &=& 1-\frac{3}{400}\left(\frac{g_A}{\pi
F}\right)^2\int^\infty_0 dpp^4F_{\pi
NN}(p^2)\nonumber\\
&\times&\left\{\frac{9}{w^3_\pi(p^2)}+\frac{6}{w^3_K(p^2)}+\frac{1}{w^3_\eta(p^2)}
\right\},\label{eq:zh} \\
 Z_s  &=& 1
-\frac{3}{400}\left(\frac{g_A}{\pi F} \right)^2\int^\infty_0
dpp^4F_{\pi
NN}(p^2)\nonumber\\
&\times&\left\{\frac{12}{w^3_K(p^2)}+\frac{4}{w^3_\eta(p^2)}
\right\}.\label{eq:zs}
\end{eqnarray}
For a meson with three-momentum $\vec p$ the meson energy is
$w_\Phi(p^2)=\sqrt{M^2_\Phi+p^2}$ with $p=|\vec p|$ and $F_{\pi
NN}(p^2)$ is the $\pi NN$ form factor normalized to  unity at zero
recoil ($\vec p=0$) :
\begin{eqnarray}
F_{\pi NN}(p^2)\!&=&\!\exp\!\left(\!\!-\frac{p^2R^2}{4}\!\right)
\!\left\{\!1\!+\!\frac{p^2R^2}{8}\!\left(\!1\!-\!\frac{5}{3g_A}\!\right)\!\right\}.
\label{eq:fpnn}
\end{eqnarray}
By adding the renormalized current quark mass term to the Dirac
equation of Eq.~(\ref{eq:dirac}) we obtain the renormalized quark
field $\psi^r$ as :
\begin{eqnarray}
\psi_i^r(x;m_i^r)=b_0u_0^r(\vec x;m_i^r)
\exp[-i{\mathcal E}_0^r(m_i^r)t],\label{eq:psir}
\end{eqnarray}
where $i$ is the flavor SU(3) index. The renormalized single quark
WF $u_0^r(\vec x;m_i^r)$ and energy ${\mathcal E}_0^r(m_i^r)$ are
related to the bare expressions $u_0(\vec x)$ and ${\mathcal E}_0$
as :
\begin{eqnarray}
u_0^r(\vec x;m_i^r)&=&u_0(\vec x)+\delta u_0(\vec x;m_i^r),\\
{\mathcal E}_0^r(m_i^r)&=&{\mathcal E}_0 + \delta {\mathcal
E}_0^r(m_i^r),
\end{eqnarray}
where
\begin{eqnarray}
\delta u_0(\vec x;m_i^r)&=& \frac{m_i^r}{2}\frac{\rho
R}{1+\frac{3}{2}\rho^2}\nonumber\\
&\times&\left(\frac{\frac{1}{2}+
\frac{21}{4}\rho^2}{1+\frac{3}{2}\rho^2}-\frac{\vec
x^2}{R^2}+\gamma^0\right)u_0(\vec x),\label{eq:du0}\\
\delta{\mathcal E}_0^r(m_i^r)&=& \gamma m_i^r.\label{eq:de0}
\end{eqnarray}
Introduction of the electromagnetic field $A_\mu$ is
accomplished by adding the kinetic energy term and by standard
minimal substitution in the Lagrangian of Eq.~(\ref{eq:efflg1})
and Eq.~(\ref{eq:efflg2}) with
\begin{eqnarray}
\partial_\mu\psi^r &\longrightarrow& D_\mu\psi^r=\partial_\mu\psi^r +
 ieQA_\mu\psi^r,\label{eq:sub1}\\
\partial_\mu\Phi_i&\longrightarrow& D_\mu\Phi_i=\partial_\mu\Phi_i +
 e\left[f_{3ij}+\frac{f_{8ij}}{\sqrt{3}}\right]A_\mu\Phi_j,\label{eq:sub2}
\end{eqnarray}
where $Q$ is the quark charge matrix and $f_{ijk}$ are the totally
antisymmetric structure constants of SU(3). The renormalized
effective Lagrangian is obtained from the original one of
Eqs.~(\ref{eq:efflg1}) and (\ref{eq:efflg2}) by replacing $\psi$ with
$\psi^r$, adding the counterterms and by standard minimal
substitution. From this we derive the electromagnetic renormalized
current operator as :
\begin{eqnarray}
j^\mu_r=j^\mu_{\psi^r}+j^\mu_\Phi+\delta j^\mu_{\psi^r}.
\label{eq:jr}
\end{eqnarray}
It contains the quark component $j^\mu_{\psi^r}$, the charged
meson component $j^\mu_{\Phi}$, and the contribution of the
counterterm $\delta j^\mu_{\psi^r}$ :
\begin{eqnarray}
j^\mu_{\psi^r}&=&\bar\psi^r\gamma^\mu Q\psi^r \nonumber\\
&=& \frac{1}{3}\left[2\bar u^r\gamma^\mu u^r-\bar
d^r\gamma^\mu d^r-\bar s^r\gamma^\mu s^r\right],\label{eq:jspi}\\
j^\mu_\Phi&=&\left[f_{3ij}+\frac{f_{8ij}}{\sqrt{3}}\right]\Phi_i
\partial^\mu\Phi_j\nonumber\\
& = &[\pi^-i\partial^\mu\pi^+ -\pi^+i\partial^\mu\pi^-\nonumber\\
&+& K^-i\partial^\mu K^+ - K^+i\partial^\mu K^-],\label{eq:jphi}
\end{eqnarray}
\begin{eqnarray}
\delta j^\mu_{\psi^r}&=&\bar\psi^r(Z-1)\gamma^\mu Q\psi^r\nonumber\\
& = &\frac{1}{3} [ 2\bar u^r(\hat Z-1)\gamma^\mu u^r-\bar
d^r(\hat Z-1)\gamma^\mu d^r\nonumber\\
&-& \bar s^r(Z_s-1)\gamma^\mu
s^r ].\label{eq:jct}
\end{eqnarray}

Following the Gell-Mann and Low theorem we define the expectation
value of an operator $\hat O$ for the renormalized PCQM by
\begin{eqnarray}
<\hat{O}>&=& {}^B\!\!<\phi_0|\sum_{n=0}^{\infty}\frac{i^n}{n!}\int
i\delta(t_1)d^4x_1...d^4x_n\nonumber\\
&\times& T[{\mathcal L}^{str}_r(x_1)...{\mathcal
L}^{str}_r(x_n)\hat O]|\phi_0>^B_c. \label{eq:op}
\end{eqnarray}
In Eq.~(\ref{eq:op}) the superscript $B$ indicates that the matrix
elements are projected on the respective baryon states, the
subscript c refers to contributions from connected graphs only and
the renormalized strong interaction Lagrangian ${\mathcal
L}^{str}_r$, which is treated as a perturbation, is defined as
\begin{eqnarray}
{\mathcal L}^{str}_r = {\mathcal L}^{str}_I+\delta{\mathcal
L}^{str},
\end{eqnarray}
where
\begin{eqnarray}
{\mathcal L}^{str}_I =
-\bar\psi^r(x)i\gamma^5\frac{\hat\Phi(x)}{F}S(r)\psi^r(x).
\label{eq:lstr}
\end{eqnarray}
$\delta{\mathcal L}^{str}$ is the strong interaction part of the
counterterms (see details in Ref.~\cite{lgf01}). We evaluate
Eq.~(\ref{eq:op}) at one loop to the order $o(1/F^2)$ using Wick's
theorem and the appropriate propagators. For the quark field we
use a Feymann propagator for a fermion in a binding potential with  
\begin{eqnarray}\label{quark_propagator}
i G_\psi(x,y) &=& <0|T\{\psi(x)\bar \psi(y)\}|0>\\ 
&=& \theta(x_0-y_0) \sum\limits_{\alpha} u_\alpha(\vec{x}) 
\bar u_\alpha(\vec{y}) e^{-i{\cal E}_\alpha (x_0-y_0)}\nonumber\\
&-& \theta(y_0-x_0) \sum\limits_{\beta} v_\beta(\vec{x})
\bar v_\beta(\vec{y}) e^{i{\cal E}_\beta (x_0-y_0)} . \nonumber
\end{eqnarray} 
Up to order of accuracy we are working in it is sufficient to use 
$G_\psi(x,y)$ instead of $G_{\psi^r}(x,y)$ where renormalized 
quark fields are used. By restricting the summation over 
intermediate quark states to the ground state we get
\begin{eqnarray}
iG_\psi(x,y) &\to& iG_\psi^{(0)}(x,y) \nonumber\\ 
&=& u_0(\vec x)\bar u_0(\vec y)e^{-i{\cal E}_0(x_0-y_0)}
\theta(x_0-y_0).\label{eq:pr}
\end{eqnarray} 
Such a truncation can be considered as an additional regularization of
the quark propagator, where in the case of SU(2)-flavor intermediate 
baryon states in loop-diagrams are restricted to $N$ and $\Delta$. 
From our previous works~\cite{lgf01}-\cite{plgfc02} we conclude that 
the use of a truncated quark propagator leads to a reasonable 
description of experimental data. 
In Ref. \cite{plgfc02} we included, for the first 
time, excited quark states in the propagator of Eq.~(\ref{quark_propagator}) 
and analyzed their influence on the matrix elements for the $N$-$\Delta$ 
transitions considered. We included the following set of excited quark 
states: the first $p$-states ($1p_{1/2}$ and $1p_{3/2}$ in the 
non-relativistic notation) and the second excited states 
($1d_{3/2}, 1d_{5/2}$ and $2s_{1/2}$). Again, we solved the Dirac equation 
analytically for the same form of the effective potential 
$V_{\rm eff}(r) = S(r) + \gamma^0 V(r)$ as was done for the ground state. 
The corresponding expressions for the wave functions of the excited quark 
states are given in the Appendix. 
In Ref. \cite{plgfc02} we demonstrated that the excited
quark states can increase the contribution of the loop diagrams  
but in comparison to the leading order (thee-quark core) 
diagram this effect was of the order of $10\%$. In the context of the 
electromagnetic properties of baryons, we also estimated the 
effect of exited states, which again is of the order of $10\%$. 
However, there are quantities (like, e.g., the charge radius of neutron) 
which are dominated by higher-order effects. Particularly, in the $SU(2)$ 
flavor limit there is no three-quark core diagram contributing to this 
quantity. Only meson-loop diagram contribute to the neutron charge radius in 
the context of the PCQM and, therefore, the effects of excited states can be 
essential. In this paper (see Sec.4) we discuss the effects of excited states 
only for the neutron charge radius. We found that these effects considerably 
improved our prediction for the neutron charge radius close to the 
experimental result.  

For the meson fields we use the free Feymann propagator for a boson with
\begin{eqnarray}
i\Delta_{ij}(x-y)&=&<0|T\{\Phi_i(x)\Phi_j(y)\}|0>\nonumber\\
&=& \delta_{ij}\int
\frac{d^4k}{(2\pi)^4i}\frac{\exp[-ik(x-y)]}{M^2_\Phi-k^2-i\epsilon}.
\label{eq:mes}
\end{eqnarray}

\section{\label{sec:emff}Electromagnetic form factors of the baryon octet}
We define the electromagnetic form factors of the baryon in the
Breit frame, where gauge invariance is fulfilled~\cite{lgf01}. In
this frame the initial momentum of the baryon is $p=(E,-\vec q/2 +
\vec{\Delta})$, the final momentum is $p^\prime=(E,\vec q/2 +
\vec{\Delta})$, and the four-momentum of the photon is $q=(0,\vec
q)$ with $p^\prime= p+q$. For identical baryons we have
$\vec\Delta = 0$. With the space-like momentum transfer squared
given as $Q^2=-q^2=\vec q^2$, we define the Sachs charge $G_E^B$
and magnetic $G_M^B$ form factors of the baryon as
\begin{eqnarray}
 <B^\prime_{s^\prime}( \frac{\vec q}{2} + \vec{\Delta})&|&J^{0}(0)|
  B_{s}(-\frac{\vec q}{2} + \vec{\Delta}) > \nonumber\\
  &=& \chi^{\dag}_{B^\prime_{s^\prime}}\chi_{B_s}G^{B}_{E}(Q^2),
  \\
  <B^\prime_{s^\prime}( \frac{\vec q}{2} + \vec{\Delta})&|&\vec J(0)|
   B_{s}(-\frac{\vec q}{2} + \vec{\Delta}) > \nonumber\\
  &=& \chi^{\dag}_{B'_{s^\prime}}\frac{i\vec \sigma_B \times \vec q}
  {m_B + m_{B^\prime}}\chi_{B_s}G^{B}_{M}(Q^2).
\end{eqnarray}
 Here, $J^0(0)$and $\vec J(0)$ are the time and space components
 of the electromagnetic current operator; $\chi_{B_s}$ and
 $\chi^{\dag}_{B^\prime_{s^\prime}}$ are the baryon spin WF in the initial
 and final states; $\vec \sigma_B$ is the baryon spin matrix.
 Electromagnetic gauge invariance both on the Lagrangian and 
 the baryon level is fulfilled in the Breit frame~\cite{lgf01}.

At zero recoil $(q^2=0)$ the Sachs form factors satisfy the
following normalization conditions:
\begin{eqnarray}
 G^{B}_{E}(0)= Q_B, \hspace*{1cm} G^{B}_{M}(0)= \mu_B,
\end{eqnarray}
where $Q_B$ and $\mu_B$ are charge and magnetic moment of the
baryon octet, respectively.

\begin{figure}[htb]
\centering
\includegraphics*[scale=1]{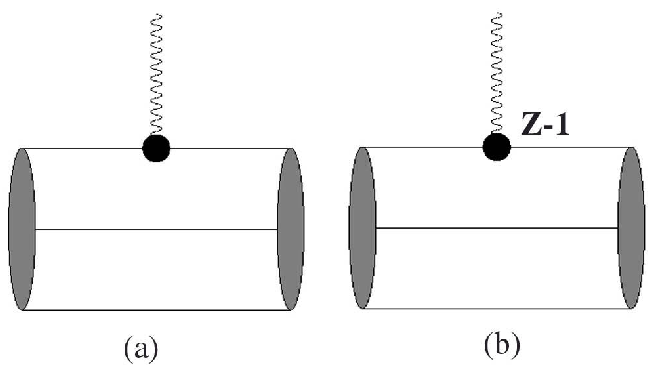} \\
\includegraphics*[scale=1]{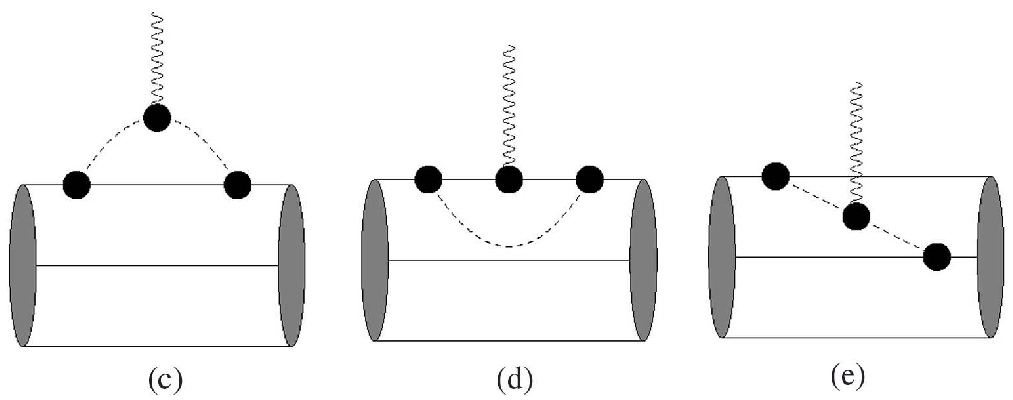}
\caption{\label{fig:diag}Diagrams contributing to the charge and
magnetic form factors of the baryon octet: three-quark diagram
(a), three-quark counterterm diagram (b), meson-cloud diagram (c),
vertex correction diagram (d), and meson-in-flight diagram (e).}
\end{figure}

The charge and magnetic radii of baryons are given by
\begin{eqnarray}
 \langle r^2_{E,M}\rangle^B=-\frac{6}{G^{B}_{E,M}(0)}
 \frac{d}{dQ^2}G^{B}_{E,M}(Q^2)\bigg|_{Q^2=0}.
\end{eqnarray}
For neutral particles $(Q_B = 0)$ the charge radius is defined by
\begin{eqnarray}
\langle
r^2_{E}\rangle^B=-6\frac{d}{dQ^2}G^{B}_{E}(Q^2)\bigg|_{Q^2=0}.
\end{eqnarray}
In the PCQM the charge and magnetic form factors of the baryon
octet are given as
\begin{eqnarray}
& &\chi^{\dag}_{s^\prime} \chi_{s} G^{B}_{E}(Q^2) \nonumber\\
&=&<\phi_{0}|\sum_{n=0}^2\frac{i^n}{n!}\!\int\delta(t) \, d^4x d^4x_1\ldots
 d^4x_n e^{-iq\cdot x} \nonumber\\
&\times&T[{\mathcal{L}}_r^{str}(x_1)\ldots{\mathcal{L}}_r^{str}(x_n)
j_r^0(x)]|\phi_{0}>_c^B,
\end{eqnarray}
\begin{eqnarray}
& &\chi^{\dag}_{s^\prime} \frac{i\vec \sigma_B \times \vec q}
  {m_B+m_{B^\prime}} \chi_{s} G^{B}_{M}(Q^2) \nonumber\\ 
&=&<\phi_{0}|\sum_{n=0}^2\frac{i^n}{n!}
  \int\!\delta(t) \, d^4x d^4x_1\ldots d^4x_n e^{-iq\cdot x} \nonumber\\
&\times&T[{\mathcal{L}}_r^{str}(x_1)\ldots{\mathcal{L}}_r^{str}(x_n)
 \vec j_r(x)]|\phi_{0}>_c^B.
\end{eqnarray}
The relevant diagrams contributing to the charge and ma\-g\-ne\-tic form
factors are indicated in Fig.~\ref{fig:diag}. In the following we
give the analytical expressions for the respective diagrams.

\begin{enumerate}
 \item Three-quark diagram (3q):
\begin{eqnarray}
 {G^{B}_{E,M}(Q^2)}\bigg|_{3q} \!\!\!&=& G^{B}_{E,M}(Q^2)\bigg|_{3q}^{\rm LO}
 \!\!\!+ G^{B}_{E,M}(Q^2)\bigg|_{3q}^{\rm NLO}\!\!\!,
\end{eqnarray}
where $G^{B}_{E,M}(Q^2)\bigg|^{\rm LO}_{3q}$ are the leading-order
(LO) terms of the three-quark diagram evaluated with the
unperturbed quark WF $u_0(\vec x);
G^{B}_{E,M}(Q^2)\bigg|^{\rm NLO}_{3q}$ is a correction due to the
modification of the quark WF $u_0(\vec x)\rightarrow 
u^r_0(\vec x;m^r_i)$ referred to as next-to-leading order (NLO):
\begin{eqnarray}
 G^{B}_{E}(Q^2)\bigg|_{3q}^{\rm LO} &=&
 a^B_{1}G^{p}_{E}(Q^2)\bigg|_{3q}^{\rm LO},
 \\
 G^{B}_{M}(Q^2)\bigg|_{3q}^{\rm LO} &=&
 b^B_{1}\frac{m_B}{m_N}G^{p}_{M}(Q^2)\bigg|_{3q}^{\rm LO},
 \\
 G^{B}_{E}(Q^2)\bigg|_{3q}^{\rm NLO} &=&
 \left(a^B_{2}+a^B_{3}\varepsilon\right)
 G^{p}_{E}(Q^2)\bigg|_{3q}^{\rm NLO},
 \\
 G^{B}_{M}(Q^2)\bigg|_{3q}^{\rm NLO} &=&
 \left(b^B_{2}+b^B_{3}\varepsilon\right)
 \frac{m_B}{m_N}G^{p}_{M}(Q^2)\bigg|_{3q}^{\rm NLO}\!\!\!,
\end{eqnarray}
where
\begin{eqnarray}
\hspace*{-.5cm}
G^{p}_{E}(Q^2)\bigg|_{3q}^{\rm LO}= 
  \exp\left(-\frac{Q^2R^2}{4}\right)
  \left(1-\frac{Q^2 R^2 \rho^2}{4(1+\frac{3}{2}\rho^2)}\right),
\end{eqnarray}
\begin{eqnarray}
\hspace*{-.5cm}  
G^{p}_{E}(Q^2)\bigg|_{3q}^{\rm NLO}&=&
   \exp\left(-\frac{Q^2R^2}{4}\right)\hat m^r
   \frac{Q^2R^3\rho}{4\left(1+\frac{3}{2}\rho^2\right)^2}\nonumber\\
  &\times&\left(\frac{1+7\rho^2+\frac{15}{4}\rho^4}{1+\frac{3}{2}\rho^2}-
  \frac{Q^2R^2}{4}\rho^2\right),
  \\
\hspace*{-.5cm}  
  G^{p}_{M}(Q^2)\bigg|_{3q}^{\rm LO}&=&
  \exp\left(-\frac{Q^2R^2}{4}\right)\frac{2m_N\rho R}{1+\frac{3}{2}\rho^2},
  \\
\hspace*{-.5cm}  
  G^{p}_{M}(Q^2)\bigg|_{3q}^{\rm NLO}&=&G^{p}_{M}(Q^2)\bigg|_{3q}^{\rm LO}
 \frac{\hat m^r R\rho}{1+\frac{3}{2}\rho^2} \nonumber\\
  &\times& \left(\frac{Q^2R^2}{4}
  -\frac{2-\frac{3}{2}\rho^2}{1+\frac{3}{2}\rho^2}\right),
  \end{eqnarray}
and $\varepsilon$ = $m^r_s$/$\hat m^r$. The constants $a^B_i$ and
$b^B_i$ are given in Table~\ref{tab:ai} and Table~\ref{tab:bi},
respectively. When using isospin symmetry we use for $m_B$, the
baryon masses, following values
\begin{eqnarray}
 m_N &=& m_p = m_n = 0.938~\mbox{GeV},\nonumber\\
 m_{\Sigma} &=& m_{\Sigma^\pm} = m_{\Sigma^0} = 1.189~\mbox{GeV},\nonumber\\
 m_{\Lambda} &=& 1.115~\mbox{GeV},\\
 m_{\Xi} &=& m_{\Xi^0} = m_{\Xi^-}= 1.321~\mbox{GeV}, \nonumber\\
 m_{\Sigma^0\Lambda} &=& \frac{1}{2}\left(m_{\Sigma}+m_{\Lambda}\right) =
 1.152~\mbox{GeV}\,.\nonumber
\end{eqnarray}
\item Three-quark counterterm (CT):
\begin{eqnarray}
 G^{B}_{E}(Q^2)\bigg|_{\rm CT} &=&
 [a^B_{2}(\hat
 Z-1) \nonumber\\
 &+& a^B_{3}(Z_s-1)]G^{p}_{E}(Q^2)\bigg|_{3q}^{\rm LO},
 \\
 G^{B}_{M}(Q^2)\bigg|_{\rm CT} &=&
 [b^B_{2}(\hat
 Z-1) \nonumber\\
 &+& b^B_{3}(Z_s-1)]\frac{m_B}{m_N}
 G^{p}_{M}(Q^2)\bigg|_{3q}^{\rm LO}.
\end{eqnarray}

\item Meson-cloud diagram (MC):
\begin{eqnarray}
G^{B}_{E}(Q^2)\bigg|_{\rm MC} &=&
 \frac{9}{400}\left(\frac{g_A}{\pi F}\right)^2
\int^\infty_0dpp^2 \int^1_{-1}dx \nonumber\\
 &\times& 
\left( p^2+p\sqrt{Q^2}x\right) {\mathcal{F}}_{\pi
NN}(p^2,Q^2,x) \nonumber\\
&\times& t^B_E(p^2,Q^2,x)\bigg|_{\rm MC},
 \\
 G^{B}_{M}(Q^2)\bigg|_{\rm MC} &=&
 \frac{3}{40}m_B\left(\frac{g_A}{\pi F}\right)^2
 \int^\infty_0dpp^4 \int^1_{-1}dx \nonumber\\
 &\times& 
(1-x^2) {\mathcal{F}}_{\pi NN}(p^2,Q^2,x) \nonumber\\
&\times& t^B_M(p^2,Q^2,x)\bigg|_{\rm MC},
\end{eqnarray}
where
\begin{eqnarray}
{\mathcal{F}}_{\pi NN}(p^2,Q^2,x) &=& 
F_{\pi NN}(p^2) \, F_{\pi NN }(p^2_+),
\nonumber\\
&&\nonumber\\
t^B_E(p^2,Q^2,x)\bigg|_{\rm MC}&=& 
a^B_4C^{11}_{\pi}(p^2,Q^2,x) \nonumber\\
&+& a^B_5C^{11}_{K}(p^2,Q^2,x),\nonumber\\
&&\nonumber\\
t^B_M(p^2,Q^2,x)\bigg|_{\rm MC}&=& 
b^B_4D^{22}_{\pi}(p^2,Q^2,x) \nonumber\\
&+& b^B_5D^{22}_{K}(p^2,Q^2,x),\\
&&\nonumber\\
D^{n_1n_2}_{\Phi}(p^2,Q^2,x)&=&
\frac{1}{w^{n_1}_{\Phi}(p^2)w^{n_2}_{\Phi}(p^2_+)}\,, \nonumber\\
&&\nonumber\\
C^{n_1n_2}_{\Phi}(p^2,Q^2,x)&=&\frac{2D^{n_1n_2}_{\Phi}(p^2,Q^2,x)}
{w^{n_1}_{\Phi}(p^2)+w^{n_2}_{\Phi}(p^2_+)}\,,\nonumber\\
&&\nonumber\\
p^2_\pm &=&p^2+Q^2 \pm 2p\sqrt{Q^2}\,.\nonumber
\end{eqnarray}

\item Vertex-correction diagram (VC):
\begin{eqnarray}
G^{B}_{E}(Q^2)\bigg|_{\rm VC} &=&
G^{p}_{E}(Q^2)\bigg|^{\rm LO}_{3q}
\frac{9}{200}\left(\frac{g_A}{\pi F}\right)^2 \nonumber\\
&\times& \int^\infty_0 dpp^4F^2_{\pi NN}(p^2)  
t^B_E(p^2)\bigg|_{\rm VC},
\\
G^{B}_{M}(Q^2)\bigg|_{\rm VC} &=&
\frac{m_B}{m_N}G^{p}_{M}(Q^2)\bigg|^{\rm LO}_{3q}
\frac{9}{200}\left(\frac{g_A}{\pi F}\right)^2 \nonumber\\
&\times& \int^\infty_0 dpp^4F^2_{\pi NN}(p^2)  
t^B_M(p^2)\bigg|_{\rm VC},
\end{eqnarray}
where
\begin{eqnarray}
 t^B_E(p^2)\bigg|_{\rm VC}&=& a^B_6W_{\pi}(p^2) +
a^B_7W_{K}(p^2) \nonumber\\
&+& a^B_8W_{\eta}(p^2),\nonumber\\
 t^B_M(p^2)\bigg|_{\rm VC}&=& b^B_6W_{\pi}(p^2) +
b^B_7W_{K}(p^2) \\
&+& b^B_8W_{\eta}(p^2),
\nonumber\\
W_{\Phi}(p^2)&=&\frac{1}{w^3_{\Phi}(p^2)} \,.\nonumber 
\end{eqnarray}
\item Meson-in-flight diagram (MF):
\begin{eqnarray}
G^{B}_{E}(Q^2)\bigg|_{\rm MF} &\equiv& 0,
\\
 G^{B}_{M}(Q^2)\bigg|_{\rm MF} &=&
 \frac{9}{100}m_B\left(\frac{g_A}{\pi F}\right)^2
\int^\infty_0dpp^4 \nonumber\\
&\times& \int^1_{-1}dx
(1-x^2) {\mathcal{F}}_{\pi NN}(p^2,Q^2,x) \nonumber\\
&\times& t^B_M(p^2,Q^2,x)\bigg|_{\rm MF},
\end{eqnarray}
where
\begin{eqnarray}
 t^B_M(p^2,Q^2,x)\bigg|_{\rm MF}&=& 
b^B_9D^{22}_{\pi}(p^2,Q^2,x) \nonumber\\
&+& b^B_{10}D^{22}_{K}(p^2,Q^2,x).
\end{eqnarray}
Due to the use of a static potential the meson-in-flight diagram 
does not contribute to the charge baryon form factors.  
\end{enumerate}
The magnetic moments $\mu_B$ of the baryon octet are given by the
expression (in units of the nucleon magneton $\mu_N$)
\begin{eqnarray}
\mu_B &=&
\mu^{\rm LO}_B\bigg[1+\delta\left(b^B_2+b^B_3\varepsilon\right) \nonumber\\
&-&\frac{1}{400}\left(\frac{g_A}{\pi F}\right)^2\int^\infty_0dpp^4
F_{\pi NN}(p^2) \nonumber\\
&\times& \left\{\frac{k^B_1}{w^3_{\pi}}+\frac{k^B_2}{w^3_K}
+\frac{k^B_3}{w^3_{\eta}}\right\}\bigg]\nonumber\\
&+&\frac{m_B}{50}\left(\frac{g_A}{\pi
F}\right)^2\int^\infty_0dpp^4
F_{\pi NN}(p^2) \nonumber\\
&\times& \left\{\frac{k^B_4}{w^4_{\pi}}
+\frac{k^B_5}{w^4_K}\right\},
\end{eqnarray}
where
\begin{eqnarray}
\mu^{\rm LO}_B=b^B_1\frac{m_B}{m_N}G_M^p(0)\bigg|^{\rm LO}_{3q}
=b^B_1\frac{2m_B\rho R}{1+\frac{3}{2}\rho^2}
\end{eqnarray}
is the leading-order contribution to the baryon magnetic moment.
The factor
\begin{eqnarray}
\delta=-\hat
m^rR\rho\frac{2-\frac{3}{2}\rho^2}{\left(1+\frac{3}{2}\rho^2\right)^2}
\end{eqnarray}
defines the NLO correction to the baryon magnetic moments due to
the modification of the quark wave function (see Eq.~(\ref{eq:du0})). 
The constants $k^B_i$ are given in Table~\ref{tab:table3}.

\section{\label{sec:result}Numerical Results}
Numerical results for the magnetic moments, charge and magnetic
radii of the baryon octet are given in Tables~\ref{tab:mu},
\ref{tab:r2ch} and \ref{tab:r2m}, respectively. The total results
for the electromagnetic properties are separated into
three parts: 1) the leading-order (3q[LO]) result due to the 
three quark core contribution; 2)  the corrections (3q[NLO + CT])
to the three quark core contribution  due to the renormalization 
of the quark WF (NLO) and the three-quark counterterm (CT) and  
3) the effects of meson loops. The meson loop contributions include  
the meson-cloud (MC), the vertex-correction (VC), and
the meson-in-flight (MF) diagrams. Experimental data are given in the 
last column of Tables. As was already mentioned, 
for a static potential the meson-in-flight diagram does not  
contribute to the baryon charge form factor. The range of our numerical 
results is due to variation of the size parameter R in the region 
0.55 - 0.65~fm. The mesonic contributions to the baryon magnetic moments
are of the order of 20 - 40 \% (except for $\Xi^-$ they contribute
only 3 \%). Hence, meson cloud corrections generate a significant
influence on baryon magnetic moments. Our results for the baryon
magnetic moments are in good agreement with the experimental data.
Mesonic contributions to the charge radii of charged baryons are
also of 20 - 40 \% (except for $\Xi^-$ where they contribute less
than 1 \%). We predict that
\begin{eqnarray}
 \left<r^2_E\right>^{\Sigma^+} > \left<r^2_E\right>^p >
\left<r^2_E\right>^{\Sigma^-} > \left<r^2_E\right>^{\Xi^-}.
\end{eqnarray}
Our result for the proton and $\Sigma^-$ charge radii squared are
in good agreement with the experimental data. In the isospin limit
the three-quark core does not contribute to the charge radii of
neutral baryons. Only the meson cloud generates a nonvanishing
value for the charge radii of these baryons. Since we restrict the
quark propagator to the ground state contribution the meson-cloud
effects give a small value for the neutron charge radius squared.
We found that the result of the neutron charge radius can be
improved by including excited states in the quark propagator. In
Table~\ref{tab:r2ch} we give a comparison of our results for the
neutron charge radius squared with the experimental value. The
value, where the quark propagator is restricted  to the ground
state, is indicated by $\left<r^2_E\right>^{n}$(GS). Contributions
from excited states (we have use $1p_{1/2}$, $1p_{3/2}$,
$1d_{3/2}$, $1d_{5/2}$ and $2s_{1/2}$) are denoted by
$\left<r^2_E\right>^{n}$(ES). Exemplified for the neutron charge
radius, we conclude that excited state contributions can also
generate sizable corrections when the LO results is vanishing. 
This result should be viewed as a first indication that excited quark state
contributions are influential in ultimately determining observables
which are dominated by loop diagrams. At this level a truncation
following the $2s_{1/2}$ state is not necessarily justified by convergence
arguments. An additional scale set by the finite size of the mesons
should be introduced to restrict the contribution of intermediate excited
quark states.
In a further effort we intend to improve our calculations to the
whole baryon octet by adding the excited states to the quark
propagator and investigating its convergence properties.
For $\Sigma^0$, $\Lambda$ and $\Xi^0$ we predict that
their charge radii squared have a positive sign and follow the
pattern
\begin{eqnarray}
\left<r^2_E\right>^{\Xi^0} > \left<r^2_E\right>^{\Sigma^0}\, , 
\left<r^2_E\right>^{\Lambda}.
\end{eqnarray}
The mesons also play a very important role for the baryon magnetic
radii where they contribute up to 50 \%. Our result for the
magnetic radius of $\Xi^-$ is quite small compared to the other's
because the meson-cloud contribution comes with a negative sign.
Results for the magnetic radii squared of the proton and neutron
are in good agreement with the experimental data. 

\begin{figure}[htb]
\includegraphics*[scale=0.4]{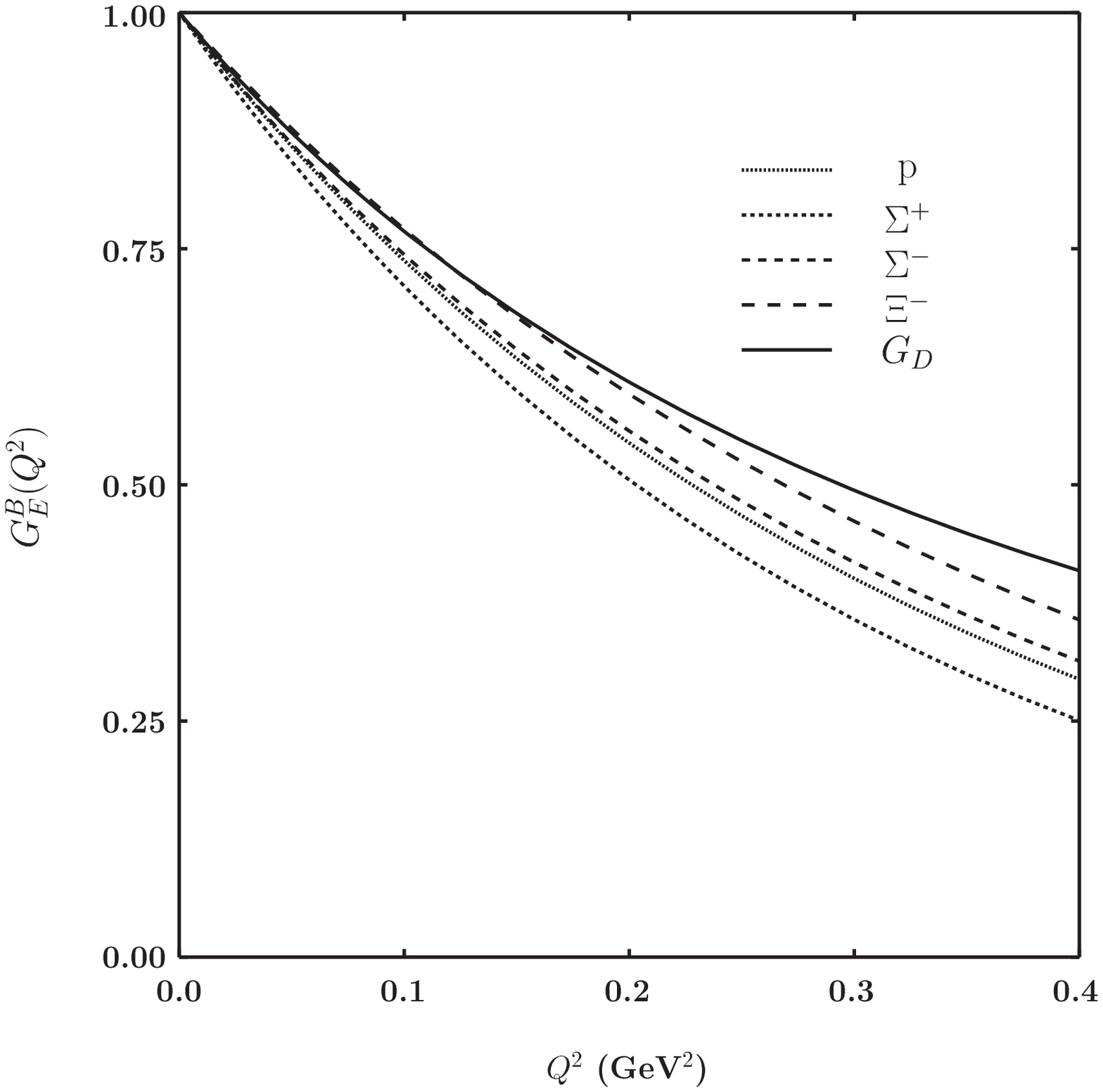}
\caption{\label{fig:ge_1}The charge form factors
$G^B_E(Q^2)$ for B = p, $\Sigma^+$, $\Sigma^-$ and $\Xi^-$ for R =
0.6 fm compared to the dipole fit $G_D(Q^2)$. For $\Sigma^-$ and
$\Xi^-$, the absolute value of $G^B_E(Q^2)$ is shown.}
\end{figure}

\begin{figure}[htb]
\includegraphics*[scale=0.4]{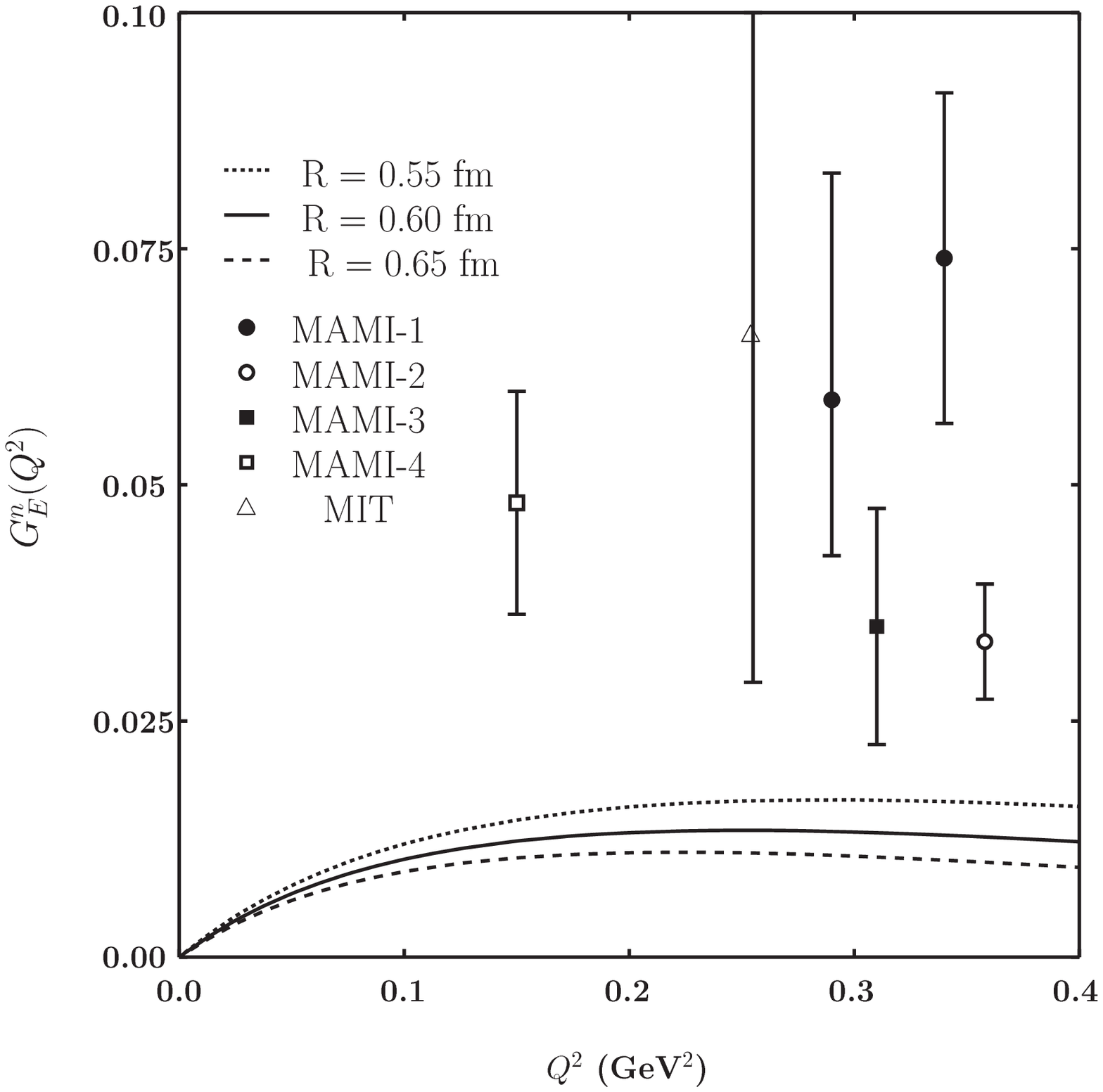}
\caption{\label{fig:ge_2}The neutron charge form
factors $G_E^n(Q^2)$ for different values of R = 0.55, 0.6, and
0.65 fm. Experimental data are taken from~\cite{mami-1}(MAMI-1),
~\cite{mami-2}(MAMI-2), ~\cite{mami-3}(MAMI-3),
~\cite{mami-4}(MAMI-4), and ~\cite{mit}(MIT).}
\end{figure}

\begin{figure}[htb]
\includegraphics*[scale=0.4]{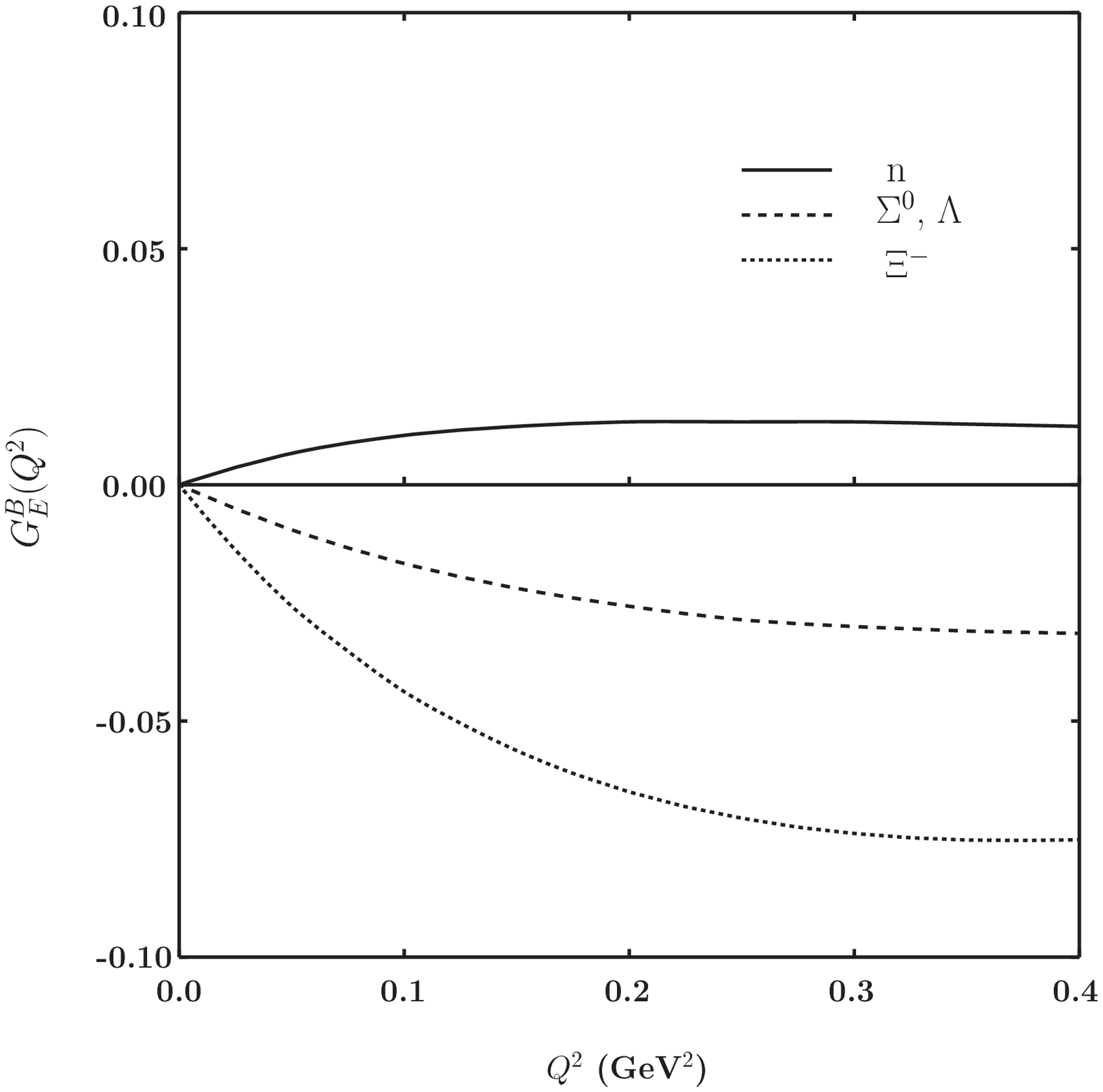}
\caption{\label{fig:ge_3} The charge form factors
$G^B_E(Q^2)$ for B = n, $\Sigma^0$, $\Lambda$ and $\Xi^0$ at R =
0.6 fm.}
\end{figure}

\begin{figure}
\includegraphics*[scale=0.4]{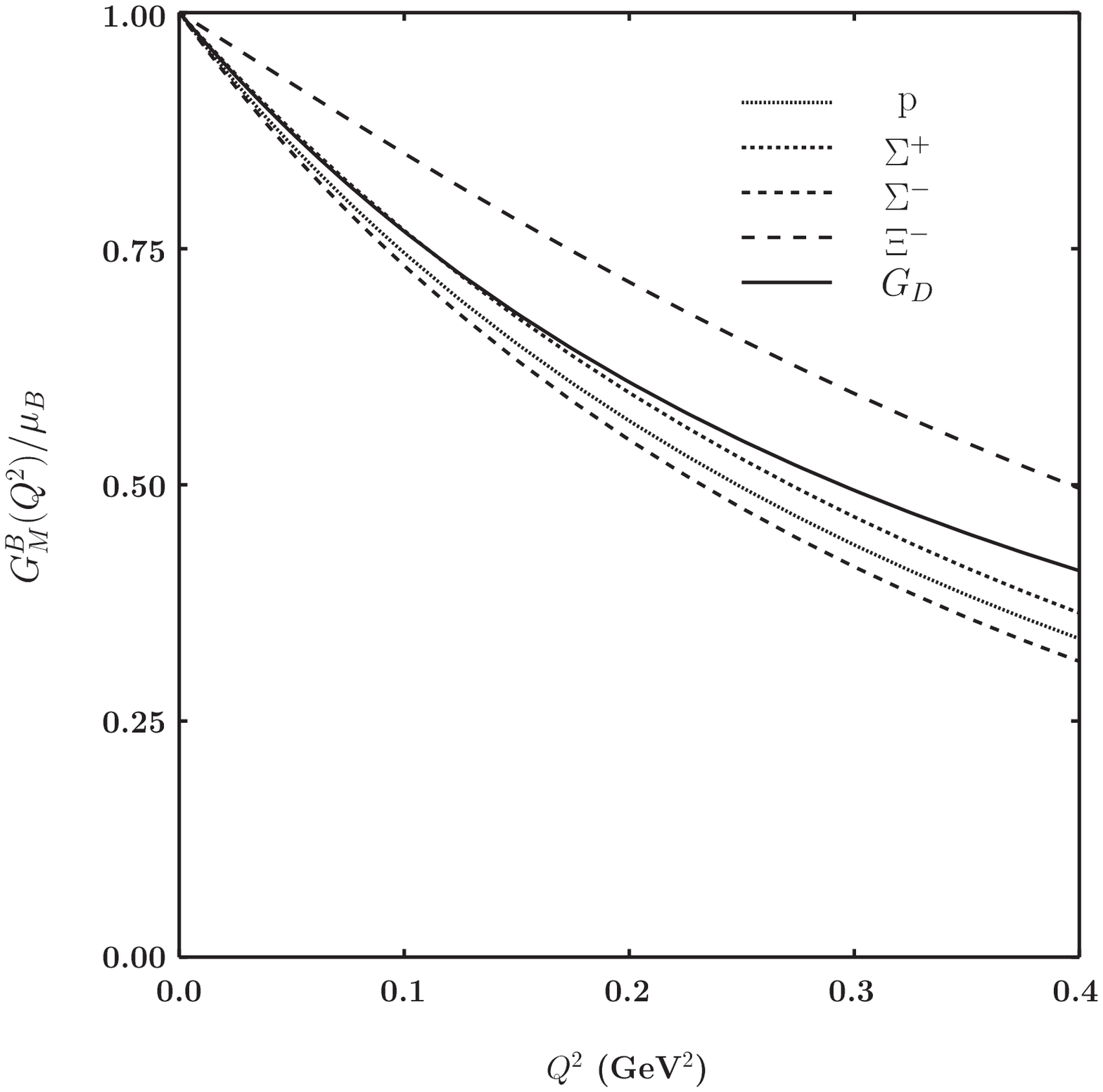}
\caption{\label{fig:gm_1}The normalized magnetic
form factors $G^B_M(Q^2)/\mu_B$ for B = p, $\Sigma^+$, $\Sigma^-$
and $\Xi^-$ at R = 0.6 fm in comparison to the dipole fit
$G_D(Q^2)$.}
\end{figure}

\begin{figure}
\includegraphics*[scale=0.4]{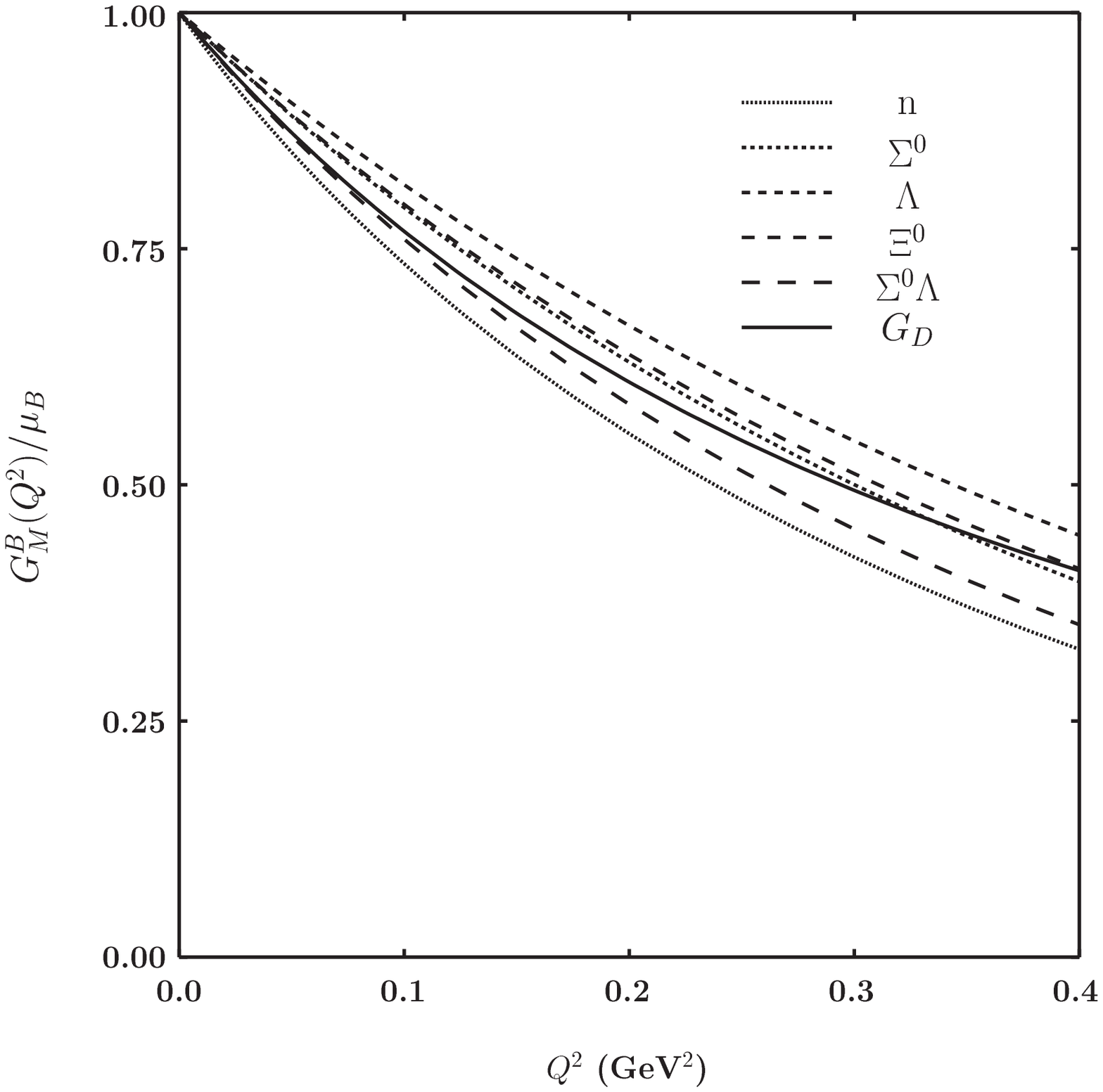}
\caption{\label{fig:gm_2} The normalized magnetic
form factors $G^B_M(Q^2)/\mu_B$ for B = n, $\Sigma^0$, $\Lambda$
and $\Xi^0$ at R = 0.6 fm as compared to the dipole fit
$G_D(Q^2)$.}
\end{figure}

The $Q^2$-dependence (up to 0.4 GeV$^2$) of the charge and magnetic form 
factors are shown in Figs.~\ref{fig:ge_1}, \ref{fig:ge_2}, \ref{fig:ge_3},
\ref{fig:gm_1} and \ref{fig:gm_2}. Due to the lack of covariance, the 
form factors can be expected to be reasonable up to 
$Q^2 < {\bf p}^2 = 0.4$ GeV$^2$, where ${\bf p}$ is the typical  
three-momentum transfer which defines the region where relativistic 
effects $\leq 10\%$ or where the following inequality 
${\bf p}^2/(4 m_N^2) < 0.1$ is fulfilled. In Fig.~\ref{fig:ge_2} 
we compare our result for the neutron charge form factor to the
experimental data varying the parameter $R$. Results are given for
the case, where the quark propagator is restricted to the ground
state. We separate the graphs for the charged and neutral baryons
by using a proper normalization and compare to the experimental
dipole fit, originally obtained for nucleon given by
\begin{eqnarray}
G_D(Q^2)=\frac{1}{(1+ Q^2/0.71~\mbox{GeV}^2)^2}.
\end{eqnarray}
There are also detailed analyses of the electromagnetic properties 
(magnetic moments, radii, form factors) of the baryon octet 
in literature. Because we are in the position to improve our formalism 
we relegate a detailed comparison to other theoretical approaches in our 
forthcoming paper. We just mention the recent 
papers~\cite{Kubis_Meissner1,Kubis_Meissner2} where a comprehensive analysis 
of baryon form factors (nucleons \cite{Kubis_Meissner1} and baryon 
octet \cite{Kubis_Meissner2}) was performed in the context of relativistic 
baryon chiral perturbation theory.  
 
\section{\label{sec:sum}Summary}
We apply the PCQM to calculate the charge and magnetic form
factors of the baryon octet up to one loop perturbation theory.
Furthermore, we analyze the magnetic moments, charge and magnetic
radii.
Since the PCQM is a static model, Lorentz covariance cannot be
fulfilled. Approximate techniques to account for Galilei invariance
and Lorentz boost effects were shown to change the tree level results
by about 10 \% \cite{luthomas}. Higher order, that is loop
contributions, are less sensitive to these corrections.
We demonstrated that meson cloud corrections play a sizable
and important role in reproducing the experimental values both for
magnetic moments and for the charge and magnetic radii. The
magnetic moments of the baryon octet can be reproduced rather
well. Also, charge and magnetic radii are explained with the PCQM,
when the LO contribution, that is the valence quarks, dominates as
soon as the LO result vanishes, meson cloud corrections which then
control the observable tend to be sensitively influenced by the
possible contributions of excited states in the loop diagrams. We
demonstrated this effect for the case of the neutron charge
radius, where inclusion of the excited states tend to improve the
model result. Further investigations, which concern the role of
excited states in calculating baryon observables are currently in
progress.

In order to improve our model we are currently pursuing following aspects. 
First, we intend to improve our calculations to the 
whole baryon octet by adding the excited states to the quark
propagator and investigating its convergence properties. Second, in
order to improve the $Q^2$-dependence of baryonic form factors we intend 
to include short-distance effects in addition to the light-meson cloud 
contributions which are important only at very low $Q^2$. 
These short-distance effects might be taken into account by the use of
low energy constants (LECs) as in chiral perturbation theory or by the use of
additional vector meson contributions as is well known to improve the 
intermediate $Q^2$-dependence. 

\begin{acknowledgement} 
This work was supported by the DFG (Grant Nos. FA67/25-3, GRK 683).
S.Cheedket and Y.Yan acknowledge the support of 
the DAAD (Grant No. a/00/27860) and Thailand Research
Fu\-nd (TRF, Grant No. RGJ PHD/00165/2541).
K.P. thanks the Development and Promotion 
of Science and Technology 
Talent Project (DPST), Thailand for financial support.  
\end{acknowledgement} 

\appendix

\section{Solutions of the Dirac equation for the effective potential}

We state here again the variational Gaussian ansatz in Eq.~(\ref{u_0})
\begin{equation}\label{Gaussian_Ansatz}
u_0({\vec x}) \, = \, N \, \exp\biggl[-\frac{{\vec x}^{\, 2}}{2R^2}\biggr]
\, \left(
\begin{array}{c}
1\\
i \rho \, {\vec \sigma}\cdot{\vec x}/R\\
\end{array}
\right)
\, \chi_s \, \chi_f \, \chi_c,
\end{equation}
This ansatz, when put back into the Dirac equation, restricts the form of the
effective potential $V_{\rm eff}(r)$ to be (note that $r=|{\vec x}|$)
\begin{equation}
V_{\rm eff}(r) = S(r) + \gamma^0 V(r)
\end{equation}
where the scalar $S(r)$ and time-like vector $V(r)$ parts are
given by
\begin{eqnarray}\label{poten}
S(r)&=& M_1 + c_1 r^2,
\nonumber \\
V(r) &=& M_2 + c_2 r^2,
\end{eqnarray}
in which $M_1$, $M_2$, $c_1$ and $c_2$ are
\begin{eqnarray}
M_1 &=& \frac{1 \, - \, 3\rho^2}{2 \, \rho R} , \hspace*{0.5cm}
M_2 = {\cal E}_0 - \frac{1 \, + \, 3\rho^2}{2 \, \rho R} , \hspace*{0.2cm} \nonumber\\
c_1 &\equiv& c_2 =  \frac{\rho}{2R^3} .
\end{eqnarray}
This specific choice of $V_{\rm eff}(r)$ from the variational Gaussian
ansatz will be used in obtaining the quark WF in any state.
The quark WF $u_{\alpha}({\vec x})$ in state $\alpha$
with eigenenergy ${\cal E}_{\alpha}$ satisfies the Dirac equation
\begin{equation}\label{dirac}
[-i {\bf \alpha} \cdot {\bf\nabla} +\beta S(r) + V(r) - {\cal E}_{\alpha}]
u_{\alpha} ({\vec x}) = 0.
\end{equation}
Due to our choice of the $V_{\rm eff}(r)$ the Dirac equation can be solved analytically and
the solutions of the Dirac spinor $u_{\alpha}({\vec x})$ to Eq. (\ref{dirac}) can
be written in the form \cite{Tegen}
\begin{equation}
u_{\alpha}({\vec x})= N_{\alpha}
\left( \begin{array}{c} g_{\alpha}(r) \\ i {\vec \sigma}
\cdot \hat {\vec x} f_{\alpha}(r) \end{array} \right) {\cal Y}_{\alpha}(\hat {\vec x})
\chi_f \chi_c .
\end{equation}
The radial functions $g_\alpha (r)$ and $f_\alpha (r)$ has the explicit form
\begin{equation}
g_{\alpha}(r) = \bigg( \frac{r}{R_{\alpha}} \bigg)^l
L^{l+1/2}_{n-1}\bigg( \frac{r^2}{R^2_{\alpha}} \bigg) e^{-\frac{r^2}{2
R^2_{\alpha}}},
\end{equation}
where for $j=l+\frac{1}{2}$
\begin{eqnarray}
f_{\alpha}(r)&=&\rho_{\alpha} \bigg(\frac{r}{R_{\alpha}}\bigg)^{l+1}
\bigg[L^{l+3/2}_{n-1}(\frac{r^2}{R^2_{\alpha}}) \nonumber\\
&+& L^{l+3/2}_{n-2}(\frac{r^2}{R^2_{\alpha}})\bigg] e^{-\frac{r^2}{2 R^2_{\alpha}}},
\end{eqnarray}
and for $j=l-\frac{1}{2}$
\begin{eqnarray}
f_{\alpha}(r)&=&-\rho_{\alpha} \bigg(\frac{r}{R_{\alpha}}\bigg)^{l-1}
\bigg[(n+l-\frac{1}{2})L^{l-1/2}_{n-1}(\frac{r^2}{R^2_{\alpha}}) \nonumber\\
&+& nL^{l-1/2}_{n}(\frac{r^2}{R^2_{\alpha}})\bigg] e^{-\frac{r^2}
{2 R^2_{\alpha}}}.
\end{eqnarray}
The label $\alpha=(nljm)$ characterizes the state with
principle quantum number $n=1,2,3,...$, orbital angular momentum $l$,
total angular momentum $j=l\pm \frac{1}{2}$ and projection $m$.
Due to the quadratic nature of the potential the radial
wave functions contain the associated
Laguerre polynomials $L^{k}_{n}(x)$ with
\begin{equation}
L^{k}_{n}(x)=\sum^{n}_{m=0} (-1)^m \frac{(n+k)!}{(n-m)!(k+m)!m!} x^m.
\end{equation}
The angular dependence ${\cal Y}_{\alpha}(\hat{\vec x}) \equiv
{\cal Y}_{lmj}(\hat{\vec x})$ is defined by
\begin{equation}
{\cal Y}_{lmj}(\hat{\vec x})=\sum_{m_l,m_s} (l m_l \frac{1}{2} m_s | j m)
Y_{l m_l}(\hat{\vec x}) \chi_{\frac{1}{2} m_s}
\end{equation}
where $Y_{l m_l}(\hat{\vec x})$ is the usual spherical harmonic.
$\chi_f$ and $\chi_c$ are the flavor and color part of the
Dirac spinor, respectively.

The coefficients $R_{\alpha}$ and $\rho_{\alpha}$ which are
belong to the $\alpha$ state are of the form
\begin{eqnarray}
R_{\alpha} &=& R(1 + \Delta {\cal E}_{\alpha} \rho R)^{-1/4},\\
\rho_{\alpha} &=& \rho \bigg( \frac{R_{\alpha}}{R} \bigg)^3 \nonumber
\end{eqnarray}
and are related to the Gaussian parameters $\rho$, $R$ of
Eq. (\ref{Gaussian_Ansatz}).
Here we define $\Delta {\cal E}_{\alpha} ={\cal E}_{\alpha}- {\cal E}_0$
to be the difference between the energy
of state $\alpha$ and the ground state.
$\Delta {\cal E}_{\alpha}$ depends on the quantum numbers $n$ and $l$
and is related to the parameters $\rho$ and $R$ by
\begin{equation}
(\Delta {\cal E}_{\alpha} + \frac{3 \rho}{R})^2
(\Delta {\cal E}_{\alpha} + \frac{1}{\rho R}) =
\frac{\rho}{R^3} (4 n +2 l -1)^2.
\end{equation}

The normalization condition is
\begin{equation}
\int\limits^\infty_0 d^3 x \,u^{\dagger}_{\alpha}({\vec x})
u_{\alpha}({\vec x}) = 1
\end{equation}
with this condition the normalization constant $N_\alpha$ is of the form
\begin{eqnarray}
N_{\alpha}&=&\bigg[ 2^{-2(n+l+1/2)} \pi^{1/2} R^3_{\alpha}
\frac{(2n+2l)!}{(n+l)!(n-1)!} \nonumber\\
&\times& \{1 + \rho^2_{\alpha}
(2n + l -\frac{1}{2})\} \bigg]^{-1/2} \,.
\end{eqnarray}

\newpage
\begin{table}
\caption{\label{tab:ai}The constants $a_i^B$ for the charge form
factors $G^{B}_{E}$ of the baryon octet.} \vspace*{0.2cm}
\renewcommand{\baselinestretch}{1.5}
\small
\begin{center}
\begin{tabular}{|c|c|c|c|c|c|c|c|c|c|c|}
\hline
 &$p$ &$n$ &$\Sigma^+$ &$\Sigma^0$ &$\Sigma^-$
 &$\Lambda$ &$\Xi^0$ &$\Xi^-$ &$\Sigma^0\Lambda$\\
\hline
$a_1$& 1   & 0   & 1    & 0    & -1   & 0    & 0    & -1   & 0 \\
$a_2$& 1   & 0   & $\frac{4}{3}$  & $\frac{1}{3}$  & -$\frac{2}{3}$
& $\frac{1}{3}$  & $\frac{2}{3}$  & -$\frac{1}{3}$ & 0 \\
$a_3$& 0   & 0   & -$\frac{1}{3}$ & -$\frac{1}{3}$ & -$\frac{1}{3}$
& -$\frac{1}{3}$ & -$\frac{2}{3}$ & -$\frac{2}{3}$ & 0 \\
$a_4$& 1   & -1  & 2    & 0    & -2   & 0    & 1    & -1   & 0 \\
$a_5$& 2   & 1   & 1    & 0    & -1   & 0    & -1   & -2   & 0 \\
$a_6$& $\frac{1}{2}$ & 1   & 0    & $\frac{1}{2}$  & 1    & $\frac{1}{2}$
& 0    & $\frac{1}{2}$  & 0 \\
$a_7$& -1  & -1  & -$\frac{1}{3}$ & -$\frac{1}{3}$ & -$\frac{1}{3}$ & -$\frac{1}{3}$
& $\frac{1}{3}$  & $\frac{1}{3}$  & 0 \\
$a_8$& $\frac{1}{6}$ & 0   & 0    & -$\frac{1}{6}$ & -$\frac{1}{3}$ & -$\frac{1}{6}$
& -$\frac{1}{3}$ & -$\frac{1}{2}$ & 0 \\
\hline
\end{tabular}
\end{center}
\end{table}

\begin{table}
\caption{\label{tab:bi}The constants $b_i^B$ for the magnetic form
factors $G^{B}_{M}$ of the baryon octet.} \vspace*{0.2cm}
\renewcommand{\baselinestretch}{1.5}
\small
\begin{center}
\begin{tabular}{|c|c|c|c|c|c|c|c|c|c|c|}
\hline
 &$p$ &$n$ &$\Sigma^+$ &$\Sigma^0$ &$\Sigma^-$
 &$\Lambda$ &$\Xi^0$ &$\Xi^-$ &$\Sigma^0\Lambda$\\
\hline
$b_1$& 1    & -$\frac{2}{3}$ & 1    & $\frac{1}{3}$  & -$\frac{1}{3}$
& -$\frac{1}{3}$ & -$\frac{2}{3}$ & -$\frac{1}{3}$ & -$\frac{\sqrt{3}}{3}$ \\
$b_2$& 1    & -$\frac{2}{3}$ & $\frac{8}{9}$  & $\frac{2}{9}$  & -$\frac{4}{9}$
& 0    & -$\frac{2}{9}$ & $\frac{1}{9}$  & -$\frac{\sqrt{3}}{3}$ \\
$b_3$& 0    & 0    & $\frac{1}{9}$  & $\frac{1}{9}$  & $\frac{1}{9}$
& -$\frac{1}{3}$ & -$\frac{4}{9}$ & -$\frac{4}{9}$ & 0 \\
$b_4$& 1    & -1   & $\frac{4}{5}$  & 0    & -$\frac{4}{5}$ & 0    & -$\frac{1}{5}$
& $\frac{1}{5}$  & -$\frac{2\sqrt{3}}{5}$ \\
$b_5$& $\frac{4}{5}$  & -$\frac{1}{5}$ & 1    & $\frac{3}{5}$  & $\frac{1}{5}$
& -$\frac{3}{5}$ & -1   & -$\frac{4}{5}$ & -$\frac{\sqrt{3}}{5}$ \\
$b_6$& $\frac{1}{18}$ & -$\frac{2}{9}$ & 0    & -$\frac{1}{9}$ & -$\frac{2}{9}$
& 0    & 0    & $\frac{1}{18}$ & -$\frac{\sqrt{3}}{18}$ \\
$b_7$& $\frac{1}{9}$  & $\frac{1}{9}$  & $\frac{5}{27}$ & $\frac{5}{27}$
& $\frac{5}{27}$ & -$\frac{1}{9}$ & -$\frac{5}{27}$ & -$\frac{5}{27}$ & 0 \\
$b_8$& -$\frac{1}{18}$& $\frac{1}{27}$ & -$\frac{2}{27}$& -$\frac{1}{27}$
& 0    & $\frac{2}{27}$ & $\frac{1}{9}$  & $\frac{5}{54}$ & $\frac{\sqrt{3}}{54}$ \\
$b_9$& 1   & -1    & 0    & 0    & 0    & 0    & 0    & 0    & -$\frac{\sqrt{3}}{3}$ \\
$b_{10}$& 0    & 0    & 1    & 1    & 1    & -1   & -1   & -1   & 0 \\
\hline
\end{tabular}
\end{center}
\end{table}

\begin{table}
\caption{\label{tab:table3} The constants $k^B_i$ for the magnetic
moment $\mu_B$ of the baryon octet.} \vspace*{0.2cm}
\renewcommand{\baselinestretch}{1.5}
\small
\begin{center}
\begin{tabular}{|c|c|c|c|c|c|c|c|c|c|}
\hline
 &$p$ &$n$ &$\Sigma^+$ &$\Sigma^0$ &$\Sigma^-$
 &$\Lambda$ &$\Xi^0$ &$\Xi^-$ &$\Sigma^0\Lambda$\\
\hline
$k_1$   & 26   & 21   & 24   & 24  & 24  & 30  & 9   & -6  & 24 \\
$k_2$   & 16   & 21   & $\frac{50}{3}$   & 14  & 22  & 0   & 25  & 32  & 18 \\
$k_3$   & 4    & 4    & $\frac{16}{3}$   & 8   & 0   & 16  & 12  & 20  & 4 \\
$k_4$   & 11   & -11  & 4    & 3   & -4  & -3  & -1  & 1   & -4$\sqrt{3}$ \\
$k_5$   & 4    & -1   & 11   & 6   & 7   & -6  & -11 & -10 & -$\sqrt{3}$ \\
\hline
\end{tabular}
\end{center}
\end{table}

\clearpage
\onecolumn

\begin{table}
\caption{\label{tab:mu}Results for the magnetic moments $\mu_B$ of
the baryon octet (in units of the nucleon magneton $\mu_N$).}

\vspace*{0.2cm}
\begin{center}
\def\arraystretch{1.75}
\hspace*{-1.7cm}
\begin{tabular}{|l|c|c|c|c|c|}
\hline
& 3q   & 3q       & Meson loops & Total & Exp~\cite{pdg02} \\
& [LO] & [NLO+CT] & [MC+VC+MF]  &       & \\
\hline $\mu_p$     & 1.80 $\pm$ 0.15 & 0.01 $\pm$ 0.03& 0.79
$\pm$0.12& 2.60 $\pm$ 0.03  & 2.793 \\
$\mu_n$            & -1.20 $\pm$ 0.10 & -0.01 $\pm$ 0.02& -0.77
$\pm$ 0.12& -1.98 $\pm$ 0.02 & -1.913 \\
$\mu_{\Sigma^+}$   & 2.28 $\pm$ 0.19 & -0.04 $\pm$ 0.04&0.51 $\pm$
0.11& 2.75 $\pm$ 0.09  & 2.458 $\pm$ 0.010\\
$\mu_{\Sigma^0}$   & 0.76 $\pm$ 0.06 & -0.05 $\pm$ 0.02 & 0.34
$\pm$ 0.07& 1.05 $\pm$ 0.01  & ---\\
$\mu_{\Sigma^-}$   & -0.76 $\pm$ 0.06 & -0.06 $\pm$ 0.01 & -0.26
$\pm$ 0.02& -1.08 $\pm$ 0.05 & -1.160 $\pm$ 0.025\\
$\mu_{\Lambda}$    & -0.71 $\pm$ 0.06 & 0.15 $\pm$ 0.04& -0.33
$\pm$ 0.09& -0.89 $\pm$ 0.03 & -0.613 $\pm$ 0.004\\
$\mu_{\Xi^0}$      & -1.69 $\pm$ 0.14 & 0.23 $\pm$ 0.09 & -0.28
$\pm$ 0.11& -1.74 $\pm$ 0.03 & -1.250 $\pm$ 0.014\\
$\mu_{\Xi^-}$      & -0.85 $\pm$ 0.07 & 0.23 $\pm$ 0.06& -0.05
$\pm$ 0.07& -0.68 $\pm$ 0.01 & -0.651$\pm$ 0.003\\
$\left|\mu_{\Sigma^0\Lambda}\right|$ & 1.28 $\pm$ 0.11 & 0.01
$\pm$ 0.02& 0.61 $\pm$ 0.09& 1.89 $\pm$ 0.01  & 1.61 $\pm$ 0.08\\
\hline
\end{tabular}
\end{center}
\end{table}

\begin{table}
\caption{\label{tab:r2ch}Results for the charge radii squared
$\left<r^2_E\right>^B$ of the baryon octet (in units of ${\rm fm}^2$).}

\vspace*{0.2cm}
\begin{center}
\def\arraystretch{1.75}
\hspace*{-2.5cm}
\begin{tabular}{|l|c|c|c|c|c|}
\hline
& 3q   & 3q       & Meson loops & Total & Exp \\
& [LO] & [NLO+CT] & [MC+VC]  &       & \\
\hline $\left<r^2_E\right>^p$   & 0.60 $\pm$ 0.10 &
0.004$\pm$0.004& 0.12$\pm$0.01& 0.72 $\pm$ 0.09
& 0.76$\pm$0.02 \cite{pdg02} \\
$\left<r^2_E\right>^{n}_{\rm GS}$         & 0   &0            & -0.043
$\pm$ 0.004& -0.043 $\pm$ 0.004 & \\
$\left<r^2_E\right>^{n}_{\rm ES}$         & 0 &0              & -0.068
$\pm$ 0.013& -0.068 $\pm$ 0.013 & \\
$\left<r^2_E\right>^{n}_{\rm Full}$ & 0&0               &
-0.111$\pm$ 0.014& -0.111 $\pm$ 0.014 & -0.116$\pm$0.002 \cite{pdg02}\\
$\left<r^2_E\right>^{\Sigma^+}$ & 0.60 $\pm$ 0.10 &
0.07$\pm$0.004& 0.14 $\pm$ 0.004& 0.81 $\pm$ 0.10    & ---\\
$\left<r^2_E\right>^{\Sigma^0}$ & 0 & 0.038$\pm$0.010 & 0.012
$\pm$ 0.010& 0.050 $\pm$ 0.010  & ---\\
$\left<r^2_E\right>^{\Sigma^-}$ & 0.60 $\pm$ 0.10  &
-0.04$\pm$0.01&0.15 $\pm$ 0.03 & 0.71$\pm$ 0.07
& 0.61 $\pm$ 0.21~\cite{sigm01} \\
$\left<r^2_E\right>^{\Lambda}$  & 0  & 0.038$\pm$0.010 & 0.012
$\pm$ 0.010& 0.050 $\pm$ 0.010  & ---\\
$\left<r^2_E\right>^{\Xi^0}$    & 0&0.07$\pm$0.02&0.07$\pm$0.02 &
 0.14  $\pm$ 0.02   & ---\\
$\left<r^2_E\right>^{\Xi^-}$    & 0.60 $\pm$ 0.10 & -0.08 $\pm$
0.03& 0.10 $\pm$ 0.03& 0.62  $\pm$ 0.07   & ---\\
$\left<r^2_E\right>^{\Sigma^0\Lambda}$ & 0&0 & 0 &0        & ---\\
\hline
\end{tabular}
\end{center}
\end{table}

\clearpage

\begin{table} \caption{\label{tab:r2m}Results for the magnetic radii
squared $ \left<r^2_M\right>^B$ of the baryon octet (in units of
${\rm fm}^2$).}

\vspace*{0.2cm}

\begin{center}
\def\arraystretch{1.5}
\hspace*{-1.7cm}
\begin{tabular}{|l|c|c|c|c|c|}
\hline
& 3q   & 3q       & Meson loops & Total & Exp \\
& [LO] & [NLO+CT] & [MC+VC+MF]  &       & \\
\hline $\left<r^2_M\right>^p$     & 0.37$\pm$0.09 &0.03$\pm$0.001&
0.34$\pm$0.02 & 0.74$\pm$ 0.07 & 0.74$\pm$0.10~\cite{simon80}  \\
$\left<r^2_M\right>^n$         & 0.33$\pm$0.08   &0.03$\pm$0.002&
0.43$\pm$0.01 & 0.79$\pm$ 0.07 & 0.76$\pm$0.02~\cite{kubon02} \\
$\left<r^2_M\right>^{\Sigma^+}$   & 0.45$\pm$0.10 &0.02$\pm$0.006
&0.17$\pm$0.02 & 0.64$\pm$ 0.08 & ---\\
$\left<r^2_M\right>^{\Sigma^0}$  & 0.39$\pm$0.10
&-0.02$\pm$0.01&0.32$\pm$0.03 & 0.69$\pm$ 0.07 & ---\\
$\left<r^2_M\right>^{\Sigma^-}$  & 0.38$\pm$0.08   &0.09$\pm$0.01&
0.31$\pm$0.01 & 0.78$\pm$ 0.07 & ---\\
$\left<r^2_M\right>^{\Lambda}$   & 0.44$\pm$0.12 &-0.14$\pm$0.06&
0.35$\pm$0.07 & 0.65$\pm$ 0.05 & --- \\
$\left<r^2_M\right>^{\Xi^0}$     & 0.52$\pm$0.12 &-0.01$\pm$0.04&
0.03$\pm$0.05 & 0.54$\pm$ 0.06 & ---\\
$\left<r^2_M\right>^{\Xi^-}$     & 0.67$\pm$0.17 &-0.31$\pm$0.12&
-0.04$\pm$0.13 & 0.32$\pm$ 0.04 & ---\\
$\left<r^2_M\right>^{\Sigma^0\Lambda}$  &
0.36$\pm$0.09&0.03$\pm$0.001
& 0.36$\pm$0.02 & 0.75$\pm$ 0.07 & ---\\
\hline
\end{tabular}
\end{center}
\end{table}

\onecolumn

\twocolumn

\begin{thebibliography}{999}
\bibitem{sigm01}I. Eschrich {\it et al.} [SELEX Collaboration], 
Phys. Lett. B \textbf{522}, (2001) 233.
\bibitem{sigm99}M. I. Adamovich {\it et al.} [WA89 Collaboration], 
Eur. Phys. J. C \textbf{8}, (1999) 59.
\bibitem{lgf01}V. E. Lyubovitskij, Th. Gutsche and A. Faessler, 
Phys. Rev. C \textbf{64}, (2001) 065203. 
\bibitem{lgfd01}
V. E. Lyubovitskij, Th. Gutsche, A. Faessler and E.G. Drukarev, 
Phys. Rev. D \textbf{63}, (2001) 054026.
\bibitem{lgfv01}
V. E. Lyubovitskij, Th. Gutsche, A. Faessler and R. Vinh Mau, 
Phys. Lett. B \textbf{520}, (2001) 204;
Phys. Rev. C \textbf{65}, (2002) 025202.
\bibitem{lgfw01}
V. E. Lyubovitskij, P. Wang, Th. Gutsche and \\ 
A. Faessler, Phys. Rev. C \textbf{66}, (2002) 055204.
\bibitem{lgff02}F. Simkovic, V.E. Lyubovitskij, Th. Gutsche, 
A. Faessler and S. Kovalenko, Phys. Lett. B \textbf{544}, (2002) 121.
\bibitem{plgfc02}
K. Pumsa-ard, V. E. Lyubovitskij, Th. Gutsche, A. Faessler and S.
Cheedket, Phys. Rev. C \textbf{68}, (2003) 015205. 
\bibitem{birse}
M. C. Birse, Prog. Part. Nucl. Phys. \textbf{25}, (1990) 1.
\bibitem{luthomas}
D. H. Lu, A. W. Thomas and A. G. Williams, 
Phys. Rev. C \textbf{57}, (1998) 2628.
\bibitem{Kubis_Meissner1}B. Kubis and U.-G. Mei\ss ner,  
Nucl. Phys. A \textbf{679}, (2001) 698.  
\bibitem{Kubis_Meissner2}B. Kubis and U.-G. Mei\ss ner, 
Eur. Phys. J. C \textbf{18}, (2001) 747.  
\bibitem{pdg02}
K. Hagiwara {\it et al.} [Particle Data Group Collaboration], 
Phys. Rev. D \textbf{66}, (2002) 010001.
\bibitem{simon80}G. G. Simon, F. Borkowski, C. Schmitt and 
V. H. Walther, Z. Naturforsch.  \textbf{35A}, (1980) 1.
\bibitem{kubon02}
G. Kubon {\it et al.}, Phys. Lett. B \textbf{524}, (2002) 26.
\bibitem{mami-1}
M. Ostrick {\it et al.}, Phys. Rev.  Lett. \textbf{83}, (1999) 276.
\bibitem{mami-2}
J. Becker {\it et al.}, Eur. Phys. J. A \textbf{6}, (1999) 329.
\bibitem{mami-3}
M. Meyerhoff {\it et al.}, Phys. Lett.  B \textbf{327}, (1994) 201.
\bibitem{mami-4}
C. Herberg {\it et al.}, Eur. Phys. J. A \textbf{5}, (1999) 131.
\bibitem{mit}
T. Eden {\it et al.}, Phys. Rev. C \textbf{50}, (1994) R1749.
\bibitem{Tegen} 
R. Tegen, R. Brockmann and W. Weise, Z. Phys. A \textbf{307}, (1982) 339.
\end{thebibliography}
\end{document}